# A Tunable Two-impurity Kondo system in an atomic point contact

Jakob Bork,[1,2] Yong-hui Zhang,[1,3] Lars Diekhöner,[2] László Borda,[4,5] Pascal Simon,[6] Johann Kroha,[5] Peter Wahl,[1] and Klaus Kern[1,7]

[1]*Max-Planck-Institut für Festkörperforschung, Heisenbergstrasse 1, D-70569 Stuttgart, Germany*

[2]*Institut for Fysik og Nanoteknologi and Interdisciplinary Nanoscience Center (iNANO), Aalborg Universitet, Skjernvej 4A, DK-9220 Aalborg, Denmark*

[3]*Department of Physics, Tsinghua University, Beijing 100084, China*

[4]*Department of Theoretical Physics, Budapest University of Technology and Economics, H-1111 Budapest, Hungary*

[5]*Physikalisches Institut, Universität Bonn, Nussallee 12, D-53115 Bonn, Germany*

[6]*Laboratoire de Physique des Solides, Université Paris-Sud, CNRS, UMR 8502, F-91405 Orsay Cedex, France*

[7]*Institut de Physique de la Matière Condensée, Ecole Polytechnique Fédérale de Lausanne (EPFL), CH-1015 Lausanne, Switzerland*

**Two magnetic atoms, one attached to the tip of a Scanning Tunneling Microscope (STM) and one adsorbed on a metal surface, each constituting a Kondo system, have been proposed as one of the simplest conceivable systems potentially exhibiting quantum critical behaviour. We have succeeded in implementing this concept experimentally for cobalt dimers clamped between an STM tip and a gold surface. Control of the tip-sample distance with sub-picometer resolution allows us to tune the interaction between the two cobalt atoms with unprecedented precision.**

**Electronic transport measurements on this two-impurity Kondo system reveal a rich physical scenario which is governed by a crossover from local Kondo screening to non-local singlet formation due to antiferromagnetic coupling as a function of separation of the cobalt atoms.**

If a magnetic impurity is introduced into a non-magnetic metallic environment, its spin interacts with the conduction electrons of the host. Below a characteristic temperature $T^0{}_K$, this leads due to the Kondo effect to a non-magnetic singlet ground state, where the spin of the magnetic impurity is completely screened [1]. As soon as two or more magnetic impurities are in proximity to each other, a competition between Kondo screening and magnetic interaction among the spins sets in and the ground state depends sensitively on their respective magnitude. Many of the peculiar properties of correlated electron materials are attributed to this competition between screening of local spins and magnetic interaction of neighboring spins. Depending on which interaction dominates the properties can vary, e.g., between antiferromagnetic ordering and heavy fermion behaviour, between being an insulator or a metal. Certain materials can be tuned continuously between the aforementioned phases through a quantum phase transition (QPT) as a function of an external control parameter, such as doping [2], pressure [3] or magnetic field [4]. The physics close to the quantum critical point is often described based on models in which the formation of dimers governs the physics [5,6]. Thus the study of a model system consisting of a dimer, as presented here, with magnetically coupled spins can serve as a test ground for these theories.

The physics of two Kondo impurities coupled by some interaction has been studied theoretically in great detail [7-13]. The relative strengths of an antiferromagnetic exchange interaction $I$ and of the single impurity Kondo temperature $T^0{}_K$ determines whether the ground state of the two-impurity system is governed by the formation of a singlet state between the two spins due to antiferromagnetic coupling or a Kondo singlet of each spin due to screening by the conduction electrons [10]. A schematic phase diagram of the two-impurity Kondo problem with antiferromagnetic coupling $I$ is displayed in fig. 1A. If the transition between the two ground states is governed by quantum fluctuations it constitutes a QPT.

Despite intense theoretical research on the two-impurity Kondo problem, an experimental realization providing access to the full phase diagram and the transition between Kondo screening and antiferromagnetism has remained elusive. Studies of semiconductor structures containing two quantum dots with a tunable coupling have been able to reveal the splitting of the Kondo resonance in the limit of strong coupling [14,15], however they exposed difficulties in tuning the two quantum dots continuously through the full phase diagram: either the single impurity regime was not accessible [15] or the splitting of the Kondo resonance, a measure for the exchange interaction, remained constant as a function of coupling in the strong coupling regime [14], which is difficult to reconcile with theory [11,12]. Also a realization with two magnetic atoms adsorbed on a metal surface has proven difficult. The limitation to a discrete set of interatomic separations imposed by the substrate lattice prevents a detailed investigation of the influence due to the exchange coupling in the interesting regime [16,17]. Only recently the non-equilibrium transport properties of the two-impurity Kondo problem have been studied theoretically for a configuration where one magnetic atom is attached to the tip of an STM and the other sits on a metal surface [18]. The coupling between the two atoms can be tuned continuously by varying the tip-sample separation, such that the system can be driven through the QPT (fig. 1A). At the QPT, an additional resonance is expected to appear in the differential conductance at zero bias whose width is only determined by the temperature of the experiment [13,18]. For the observability of the QPT the properties of the two Kondo impurities are not required to be identical, which facilitates an experimental realization. We have implemented this model system with two cobalt atoms as the Kondo impurities. The measurement setup is depicted in fig. 1B.

Following sample preparation as described in the Supporting Online Material, in STM images measured at ~8K we find single cobalt adatoms on the Au(111) surface (fig. 2A), which can be identified by their characteristic Kondo resonance.[19] In tunneling spectra, the Kondo resonance is detected as a Fano line shape (eq. S1) due to interference between electrons tunneling into the conduction band of the substrate and into the Kondo state (fig. 2E). Depending on the relative strengths of the two channels, the line shape of the Kondo resonance can be a dip, an asymmetric line shape or a peak[19]. We have attached single cobalt atoms to the STM tip by positioning the tip on top of an adatom and applying a voltage pulse (fig. 2B)[20]. After application of a successful pulse the cobalt atom disappears from the sample surface (see fig. 2C) and the spectrum taken over a clean spot of the gold surface exhibits a similar resonance compared to the one found for cobalt atoms on the sample surface (see fig. 2F). This suggests that the atom at the tip has a similar environment as on the surface[22]. Though the STM tip is made from a chemically etched tungsten wire, its apex is covered with gold due to indentation of the tip into the sample which we perform for tip conditioning. As the precise environment of the adatom on the tip cannot be controlled well, there is some scattering in the appearance of the tip Kondo resonance for different tip preparations which influences details of the spectra (see fig. S4 for statistics and typical spectra after pick-up). When the tip with cobalt atom is positioned on top of a cobalt atom on the surface, the tunneling spectrum shows a superposition of the Kondo resonances of tip and sample as is apparent from the larger amplitude of the resonance in the tunneling spectrum (see fig. 2G).

While approaching the two cobalt atoms towards each other, the junction makes a transition from the tunneling to the point contact regime (fig. 3A). Reversible formation of point contacts between an STM tip and a single atom on a metal surface has been shown previously both for non-magnetic [21,23] and magnetic adatoms [24,25]. Spectra acquired in the tunneling regime for $z > 0$ Å (see fig. 2G, fig. 3B) show a resonance similar to the one found for a single cobalt atom in the junction (compare figs. 2E, F) however with a larger amplitude as it is a superposition of the Kondo resonances of tip and sample. Shape and width of the feature change only little as long as the tip-sample distance stays comparatively large ($z > 2$ Å). Once the two atoms are moved within 2 Å of point contact the width of the resonance is reduced. Close to the transition to the point contact regime, the resonance changes its shape from a dip to a *peak* of similar width (see fig. 3B for $z \sim -30$ pm). On pushing the tip further in, the peak splits into two resonances at symmetric positions with respect to zero bias. The splitting increases with decreasing tip-sample distance (fig. 3B). We note that all data shown in this work is obtained from reversible, non-destructive measurements, where STM images taken before and after formation of the point contact show no changes of the tip or the cobalt atom itself. The conductance traces during approach and withdrawal are apart from some vertical drift coincident with each other. Occasionally we observe lateral hops of the cobalt atom which can however be directly detected in the conductance trace as well as from the STM images taken before and after the approach curve is acquired.

With only one cobalt atom between tip and surface, neither do we see a reduction of the width of the resonance (consistent with refs. 24, 25), nor a change in the line shape from dip to peak nor a splitting (compare fig. S2). Thus these effects are related to the presence of two cobalt atoms in the tunneling junction. This renders inelastic tunneling due to vibrational excitations[26] as a possible source for the conductance curve unlikely. Effects due to contamination, e.g. with hydrogen, typically lead to characteristic spectra which are distinctly different from the conductance curves shown in figs. 2 and 3[27], furthermore in this case spectra should not depend on whether there are one or two cobalt atoms in the junction.

We can identify two regimes from the conductance traces: a weak-coupling regime, in which only one resonance is observed and a strong-coupling regime, where we find a split resonance. To understand the evolution of the spectra we have performed Numerical Renormalization Group (NRG) calculations. NRG is a numerically exact and non-perturbative method to describe the physics of quantum impurities[28,29]. In fig. 3C, one diagonal element of the T-matrix is plotted, which is proportional to the local density of states (LDOS) at the two impurity sites. The parameters of the calculation are defined in fig. 1B, different tip-sample distances are modeled by different couplings $\Delta$ between the two electrodes. We find qualitative agreement between our experiment and NRG calculations concerning the evolution from a single peak towards a split resonance.

The splitting of the resonance directly reflects the exchange interaction between the spins of the two cobalt atoms[11,12,17] which leads to a splitting of states with even and odd parity. We can extract its dependence on the vertical distance between the two electrodes by fitting a sum of two Fano line shapes to the experimental data[17]. The resulting splitting is plotted as a function of vertical tip position in fig. 4A. It rises within 20 pm from 3 to 9meV. For compact cobalt dimers on Au(111) and Cu(100), the Kondo resonance has been observed to be completely suppressed[16,17] , which has been attributed to a reduced coupling of the dimer magnetic moment to the conduction band[16] and strong magnetic interaction significantly larger than the Kondo energy[17]. We do not observe this suppression of the Kondo resonance. One possible reason for this apparently large difference in the interaction can be the relative spatial orientation of the $d$-orbitals which contribute to the Kondo effect. In case of a $d_{xy}$ or $d_{x2\text{-}y2}$ orbital (with x and y parallel to the surface), the direct interaction in the vertical direction can be significantly smaller than in the lateral direction.

To quantitatively analyze the behaviour of the resonance in the weak coupling regime, we have fitted single Fano resonances to the spectra. The width of the resonance is reduced by up to ~ 70% (see fig. 4B) when the two cobalt atoms are brought close to each other.

The NRG calculation does not show a significant change in the width of the resonance at weak coupling. However as measurements with only one cobalt atom in the junction do not show a similar reduction of the width, it must be related to the presence of two cobalt atoms.

We can rationalize the reduction by the superposition of the Kondo resonances of tip and sample. In the spectra, only one resonance is observed because the resonances of the two cobalt atoms are both at the Fermi level (see fig. 4C for an illustration). To assess the influence of their superposition on the overall spectrum, we have performed a model calculation assuming that the bare tip and the sample spectra can be described by Fano resonances (as found when measuring with only one cobalt atom in the junction, see fig. 2E, F; Ref. 19). We assume that only the line shape of the individual resonances changes from a dip to a peak while approaching the two cobalt atoms towards each other, keeping the parameters of the Kondo resonances (i.e. width, amplitude and position) constant as found in the NRG calculation. With this model, we can recover the behaviour of the tunneling spectra and of the overall width of the resonance. A change in the line shape of the Fano resonance has been predicted to occur if the relative magnitude of the couplings of the tip to the conduction band and the Kondo state, respectively, changes[30]. This can occur if the vacuum tail of the conduction band and the Kondo state decays on different length scales. The observation of Fano resonances in the contact regime is consistent with earlier work with single cobalt atoms contacted by an STM tip[24,25] and theoretical predictions[33,34], however the interpretation in terms of different transport channels is less clear. The modeled spectra are shown in Fig 4D and demonstrate that the overall shape of the measured spectra (Fig 3B) is well reproduced. We have plotted the single overall width extracted from fitting a Fano resonance to the calculated spectra (for details see S2B) on top of the width extracted from the experimental data in fig. 4B. The agreement is very good apart from deviations close to the transition to point contact. These could be due to mechanical relaxation effects of tip and sample, which are especially strong at the transition. The assumed increase of the single impurity line shape parameter $q$ can be attributed to an increase in coupling between the $d$-orbitals of the two cobalt atoms thus opening an additional direct channel contributing to the current[19,30]. This is supported by an increased conductance in point contact which we observe for junctions containing two cobalt atoms compared to junctions with only one cobalt atom or only gold atoms (see fig. S3). Though the interaction between the conduction band of one electrode and the $d$-orbital of the cobalt atom on the opposite electrode might naively be expected to be stronger than direct interaction of the $d$-orbitals, it appears to be rather negligible as otherwise for a gold tip approaching towards a cobalt atom a change in the line shape should be observed. Alternatively the reduction of the width could be explained as being due to a weak antiferromagnetic interaction which can reduce the Kondo temperature as pointed out, e.g., in Ref. 5. Since we cannot resolve the single Kondo resonances but rather observe the superposition of the two in our spectra, we cannot rule out this possibility. However consistency with the NRG calculation as well as the increased conductance in point contact support our modeling of the reduction of the width as being due to the change in the line shape.

We can clearly identify the two ground states expected for the cobalt dimer from the two-impurity Kondo problem. The resulting phase diagram is depicted in fig. 5. Far out in the tunneling regime, the two cobalt atoms behave as weakly interacting Kondo impurities, each with its own Kondo resonance. Approaching the two cobalt atoms towards each other, the interaction rises and the resonance splits due to exchange interaction. Further reduction of the tip-sample distance increases the interaction and hence the separation of the two peaks.

Our data show a splitting of the Kondo resonance already for exchange interactions as small as 3 meV, significantly smaller than the characteristic energy scale of Kondo screening of about $k_B T_K^0 \sim 6.5$ meV and the predicted critical coupling, beyond which a sudden splitting of the Kondo resonance is expected, of $I^* = 2 k_B T_K^0 \sim 13$ meV [7,9-12]. Both, the appearance of the splitting at couplings smaller than the critical coupling as well as the shape of the resonances on entering the antiferromagnetic regime are fully consistent with the NRG calculations shown in fig. 3C. For much stronger exchange interactions, the spectra are expected to be governed by inelastic spin excitations[31] breaking the singlet rather than by a split Kondo resonance.

Instead of showing a QPT, the cobalt dimer undergoes a crossover between the two regimes, Kondo screening and antiferromagnetism. By comparison with the NRG calculations, the crossover can be attributed to the strong coupling between the electrodes. For the system to reveal quantum critical behaviour, the coupling between the electrodes needs to be small compared to the single impurity Kondo temperature [7,18]. One way to come closer to a system exhibiting critical behaviour is to use atoms as Kondo systems which exhibit a significantly lower Kondo temperature while still having substantial coupling to the electrodes (by which the magnetic coupling is mediated). This can be achieved for Kondo systems which have a Coulomb repulsion about a factor of two higher than cobalt atoms, possible candidates could be rare earth atoms[35]. Another way would be to increase the direct coupling between the orbitals which form the Kondo state. Direct charge coupling also renders the QCP unstable[36], so it will only be observable for the right balance of direct hopping and exchange interaction[37].

The atomically precise engineering of a two-atom Kondo system provides a unique playground to probe electron correlations, complementing studies on semiconductor quantum dots. Our experiment provides a model system for studying the competition between electron correlation and local moment antiferromagnetism which eventually might allow studying quantum criticality. It also opens up new perspectives for atomic scale magnetometry. The tip Kondo system can be used as a probe for magnetic interactions, yielding quantitative information about the magnetic interaction. This offers the opportunity to study magnetic properties of atoms, magnetic molecules, clusters and islands with the high lateral resolution of STM.

1. Hewson, A.C. The Kondo Problem to Heavy Fermions (Cambridge University Press, 1993).

2. v. Löhneysen, H., Pietrus, T., Portisch, G., Schlager, H.G., Schröder, A., Sieck, M. and Trappmann, T. Non-fermi-liquid behavior in a heavy-fermion alloy at a magnetic instability. Phys. Rev. Lett. **72**, 3262–3265 (1994).

3. Mathur, N.D., Grosche, F.M., Julian, S.R., Walker, I.R., Freye, D.M., Haselwimmer, R.K.W., and Lonzarich, G.G. Magnetically mediated superconductivity in heavy fermion compounds. Nature **394**, 39–43 (1998).

4. Grigera, S.A., Perry, R.S., Schofield, A.J., Chiao, M., Julian, S.R., Lonzarich, G.G., Ikeda, S.I., Maeno, Y., Millis, A.J., and Mackenzie, A.P., Magnetic field-tuned quantum criticality in the metallic ruthenate $Sr_3Ru_2O_7$. Science **294**, 329–332 (2001).

5. Klein, M., Nuber, A., Reinert, F., Kroha, J., Stockert, O., and v. Löhneysen, H. Signature of quantum criticality in photoemission spectroscopy. Phys. Rev. Lett. **101**, 266404 (2008).

6.  Sachdev, S. Quantum criticality: Competing ground states in low dimensions. Science **288**, 475–480 (2000).

7.  Georges, A. and Meir, Y. Electronic correlations in transport through coupled quantum dots. Phys. Rev. Lett. **82**, 3508–3511 (1999).

8.  Jayaprakash, C., Krishna-murthy, H.R., and Wilkins, J.W. Two-impurity Kondo problem. Phys. Rev. Lett. **47**, 737–740 (1981).

9.  Jones, B.A. and Varma, C.M. Study of two magnetic impurities in a fermi gas. Phys. Rev. Lett. **58**, 843–846 (1987).

10. Jones, B.A., Varma, C.M., and Wilkins, J.W. Low-temperature properties of the two-impurity Kondo hamiltonian. Phys. Rev. Lett. **61,** 125–128 (1988).

11. López, R., Aguado, R., and Platero, G. Nonequilibrium transport through double quantum dots: Kondo effect versus antiferromagnetic coupling. Phys. Rev. Lett. **89**, 136802 (2002).

12. Simon, P., López, R., and Oreg, Y. Ruderman-Kittel-Kasuya-Yosida and magnetic-field interactions in coupled Kondo quantum dots. Phys. Rev. Lett. **94**, 086602 (2005).

13. De Leo, L. and Fabrizio, M. Spectral properties of a two-orbital Anderson impurity model across a non-Fermi-liquid fixed point, Phys. Rev. B **69**, 245114 (2004).

14. Craig, N.J., Taylor, J.M., Lester, E.A., Marcus, C.M., Hanson, M.P., and Gossard, A.C. Tunable nonlocal spin control in a coupled-quantum dot system. Science **304**, 565–567 (2004).

15. Jeong, H., Chang, A.M., and Melloch, M.R. Kondo effect in an artificial quantum dot molecule. Science **293,** 2221–2223 (2001).

16. Chen, W., Jamneala, T., Madhavan, V., and Crommie, M.F. Disappearance of the Kondo resonance for atomically fabricated cobalt dimers. Phys. Rev. B **60**, R8529–R8532 (1999).

17. Wahl, P., Simon, P., Diekhöner, L., Stepanyuk, V.S., Bruno, P., Schneider, M.A., and Kern, K. Exchange interaction between single magnetic adatoms. Phys. Rev. Lett. **98**, 056601 (2007).

18. Sela, E. and Affleck, I. Nonequilibrium transport through double quantum dots: Exact results near a quantum critical point. Phys. Rev. Lett **102**, 047201 (2009).

19. Madhavan, V., Chen, W., Jamneala, T., Crommie, M.F., and Wingreen, N.S. Tunneling into a single magnetic atom: Spectroscopic evidence of the Kondo resonance. Science **280**, 567–569 (1998).

20. Eigler, D.M., Lutz, C.P., and Rudge, W.E. An atomic switch realized with the scanning tunneling microscope. Nature **352,** 600–603 (1991).

21. Limot, L., Kröger, J., Berndt, R., Garcia-Lekue, A., and Hofer, W.A. Atom transfer and single-adatom contacts. Phys. Rev. Lett. **94**, 126102 (2005).

22. Madhavan, V., Chen, W., Jamneala, T., Crommie, M.F., and Wingreen, N.S. Local spectroscopy of a Kondo impurity: Co on Au(111). Phys. Rev. B **64**, 165412 (2001).

23. Yazdani, A., Eigler, D.M., and Lang, N.D. Off-resonance conduction through atomic wires. Nature **272**, 1921–1924 (1996).

24. Néel, N., Kröger, J., Limot, L., Palotas, K., Hofer, W.A., and Berndt, R. Conductance and Kondo effect in a controlled single-atom contact. Phys. Rev. Lett. **98**, 016801 (2007).

25. Vitali, L., Ohmann, R., Stepanow, S., Gambardella, P., Tao, K., Huang, R., Stepanyuk, V.S., Bruno, P., and Kern, K. Kondo effect in single atom contacts: The importance of the atomic geometry. Phys. Rev. Lett. **101,** 216802 (2008).

26. Stipe, B.C., Rezaei, M.A., and Ho, W. Single-Molecule Vibrational Spectroscopy and Microscopy. Science **280**, 1732-1735 (1998).

27. Gupta, J.A., Lutz, C.P., Heinrich, A.J., and Eigler, D.M. Strongly coverage-dependent excitations of adsorbed molecular hydrogen. Phys. Rev. B **71**, 115416 (2005).

28. Wilson, K.G. The renormalization group: Critical phenomena and the Kondo problem. Rev. Mod. Phys. **47,** 773–840 (1975).

29. Bulla, R., Costi, T.A., and Pruschke, T. Numerical renormalization group method for quantum impurity systems. Rev. Mod. Phys. **80**, 395–450 (2008).

30. Plihal, M. and Gadzuk, J.W. Nonequilibrium theory of scanning tunneling spectroscopy via adsorbate resonances: Nonmagnetic and Kondo impurities. Phys. Rev. B **63**, 085404 (2001).

31. A.J. Heinrich, J.A. Gupta, C.P. Lutz, and D.M. Eigler, Single-Atom Spin-Flip Spectroscopy, Science **306**, 466 (2004).

32. Though the coupling is expected to be exponentially decaying for larger distances z, in point contact the relation between distance z and actual distance between the cobalt atoms (or electrodes) is not clear.

33. Lucignano, P., Mazzarello, R., Smogunov, A. Fabrizio, M., and Tosatti, E. Kondo conductance in an atomic nanocontact from first principles. Nat. Mat. **8**, 563 (2009).

34. D. Jacob, K. Haule and G. Kotliar, Kondo Effect and Conductance of Nanocontacts with Magnetic Impurities. Phys. Rev. Lett. **103**, 016803 (2009).

35. van der Marel, D. and Sawatzky, G.A. Electron-electron interaction and localization in d and f transition metals, Phys. Rev. B **37**, 10674 (1988).

36. Zaránd, G., Chung, C.-H., Simon, P., and Vojta, M. Quantum Criticality in a Double-Quantum-Dot System. Phys. Rev. Lett. **97**, 166802 (2006).

37. Malecki, J., Sela, E., and Affleck, I., The prospect for observing the quantum critical point in double quantum dot systems. arxiv:Cond-mat/1009.0860 (2009).

We are indebted to I. Affleck for discussion stimulating our research. Further, we acknowledge discussions with M. Fabrizio and D. Jacob.

Correspondence and requests for materials should be addressed to P.W. (wahl@fkf.mpg.de).

**Fig. 1.** Phase diagram and experimental realization of the two-impurity Kondo Problem. (**A**) Schematic phase diagram of the two-impurity Kondo model showing the relevant energy scales and the ground states as a function of coupling between the spins: KS - Kondo screening, where the single-impurity Kondo screening dominates, the Kondo temperature $T^0_K$ is the characteristic energy scale of a single Kondo impurity, the coupled system has a characteristic energy scale $T^*$; AF - local moment antiferromagnetism dominated by the magnetic coupling between the two spins, the dashed line indicates the expected behaviour of the Néel temperature $T_N$ of the two impurities; the vertical dotted line at the critical coupling $I^*$ indicates the position where the quantum phase transition (QPT) would occur. The behaviour of $T_N$ and $T^*$ close to the QPT is not clear (indicated by the dotted lines). (**B**), illustration of our measurement setup with one cobalt atom on the STM tip and one on the surface. The hybridization $V_{s,t}$ between the cobalt atoms and their respective electrodes leads to Kondo screening of the spins of the cobalt atoms. The coupling $\Delta$ between tip and sample results in an antiferromagnetic interaction $I$ between the two spins. The strength of the interaction is varied by changing the distance between the atoms. The couplings $V_{s,t}$ and $\Delta$ are the main input parameters of the NRG calculations, whereas $I$ is a consequence of these couplings.

**Fig. 2.** Preparation of the tip Kondo system. (**A**) 3D representation of an STM image of a Au(111) surface taken at ~8K with three cobalt adatoms taken with a spectroscopically featureless tip, (**B**) by applying a voltage pulse after approaching the tip towards the adatom, one cobalt atom is transferred to the tip – which can be seen in the current trace acquired during the voltage pulse as a sudden change (solid blue line refers to the right vertical axis, dashed black line to the left one), (**C**) the atom which has been picked up has disappeared from the surface. (**D**) The spectrum of the tip recorded prior to picking up an atom on the clean gold surface, which is apart from a linear background featureless. (**E**) Spectra taken on cobalt adatoms show a Kondo resonance, the Kondo temperature $T^0_K$ is determined from the half-width of the resonance [19]. (**F**) After picking up the cobalt atom, the differential conductance spectrum taken on a clean spot of the surface shows a Kondo resonance, too. (**G**) Spectra taken with a tip with cobalt atom attached on a second cobalt atom on the surface now show the two Kondo resonances superimposed, as is seen from the amplitude of the resonance.

**Fig. 3.** Measurements on two-impurity Kondo system in tunneling and transport. (**A**) Approach curve acquired while moving the tip with a cobalt atom attached towards a cobalt atom on the surface. Shown is the conductance in units of the spin-degenerate quantum of conductance $G_0$=12.9kΩ as a function of tip-sample distance at fixed bias voltage ($V_{bias}$ = 40mV); the transition from tunneling regime (blue) to point contact regime (green) can be clearly seen. (**B**) Spectra as a function of tip-sample distance (as indicated on the right) in the tunneling regime (blue) and in the point contact regime (green). (**C**) NRG calculation for the T-matrix, which is proportional to the LDOS, showing the transition from the Kondo resonance to a split peak on increasing the coupling Δ between the electrodes. Spectra in B and C are shifted vertically.

**Fig. 4.** Interaction effects between the cobalt atoms (**A**) Splitting of the peaks seen in fig. 3B as a function of tip-sample distance. Black crosses have been obtained from NRG calculations in fig. 3C, with the assumption $z \sim \log\Delta$ [32] (solid line is a linear fit to the NRG calculations as a guide to the eye). (**B**) width of the resonance extracted from tunneling spectra as shown in fig. 3B as a function of tip-sample distance, circles and triangles indicate separate sets of measurements done after different sample preparations with different tips, colors have the same meaning as in fig. 3A. The solid curve shows the calculated width of the resonance, where the line shape parameter $q$ has been assumed to depend on the tip-sample distance $z$ as $q \sim \exp(-z/\kappa_d)$. (**C**) Superposition of the Kondo resonances of tip and sample (black lines) leads to an overall tunneling spectrum (red line) as in fig. 3B (details of calculation in section S2B of the supplementary material); (**D**) change of the overall spectrum

as the line shape of the single impurity resonance changes from a dip to a peak while the tip approaches the sample.

**Fig. 5.** Schematic phase diagram emerging from the experiment. The quantum phase transition at the critical coupling is suppressed due to the strong coupling between the conduction electrons at the tip and the sample, instead the system exhibits a broad crossover region between the Kondo screened and the antiferromagnetically coupled regimes. Red and blue lines indicate the experimentally accessible lines of the diagram (solid red line extracted from data shown in fig. 4B assuming that the exchange interaction is proportional to the splitting, dashed red line indicates data shown in the fig. S5c). The splitting is observed well below the critical coupling $I^*$. Symbols and Abbreviations as in fig. 1A.

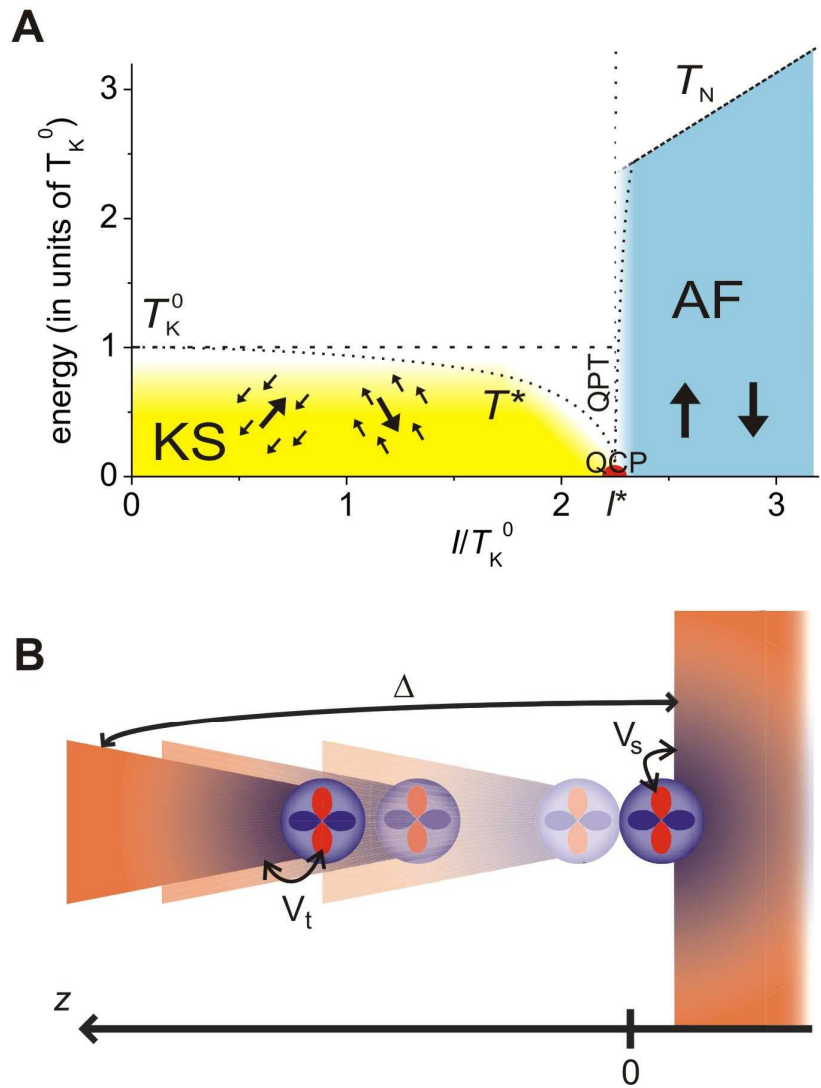

Fig. 1

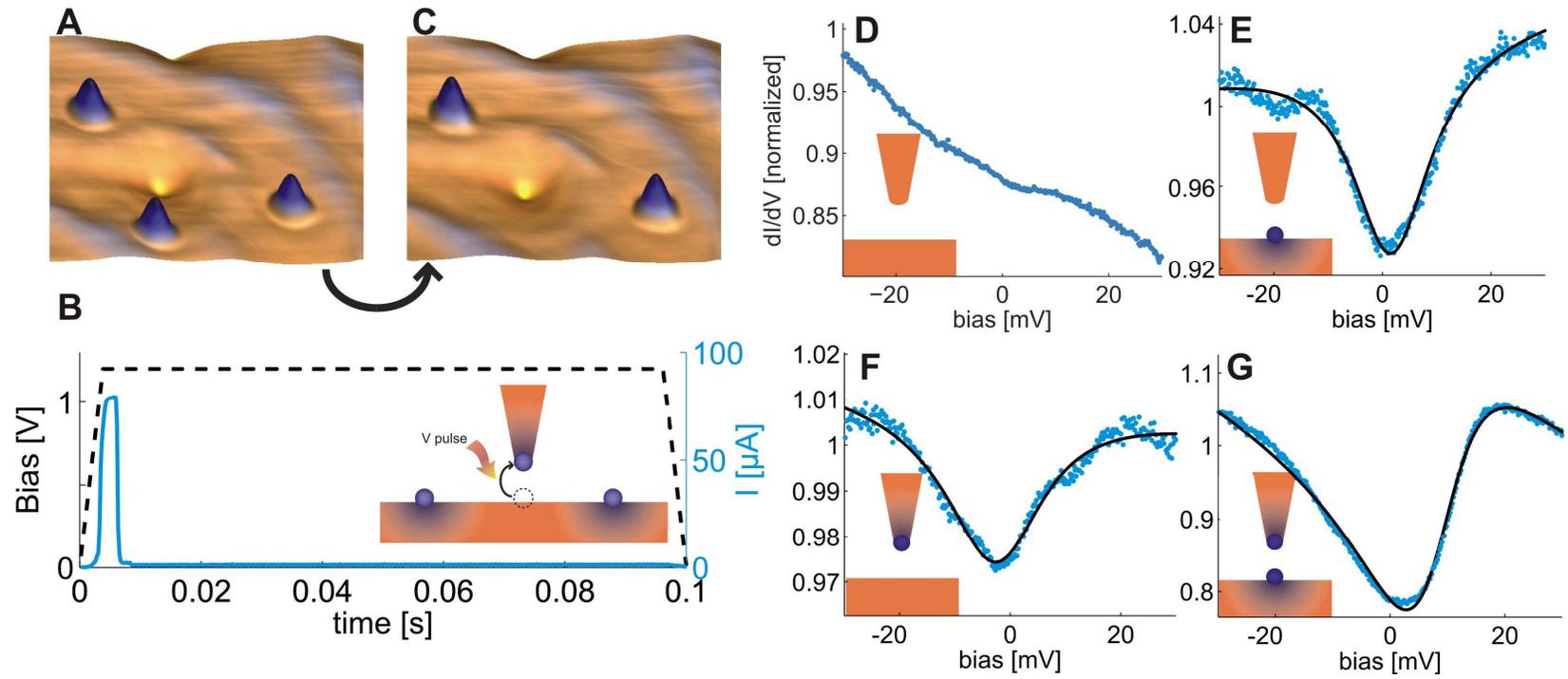

Fig. 2

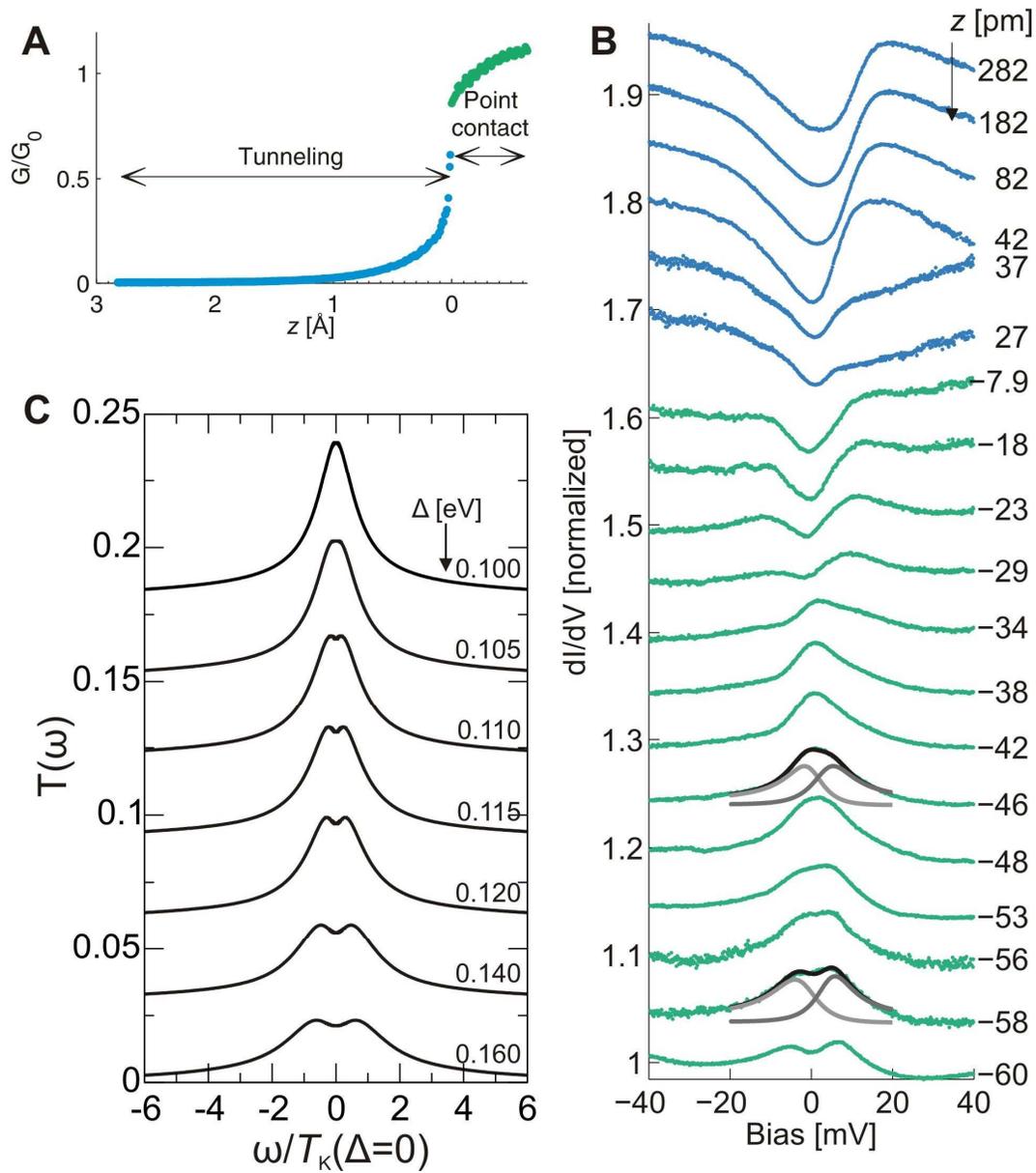

Fig. 3

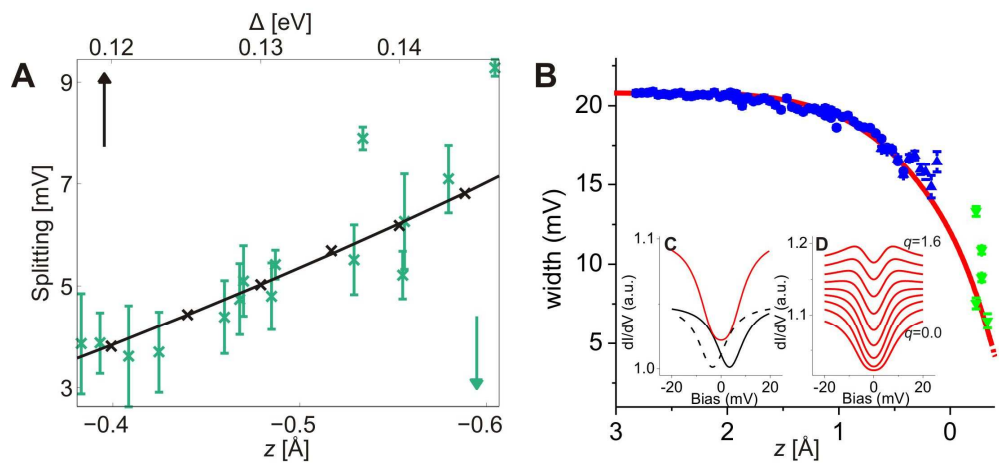

Fig. 4

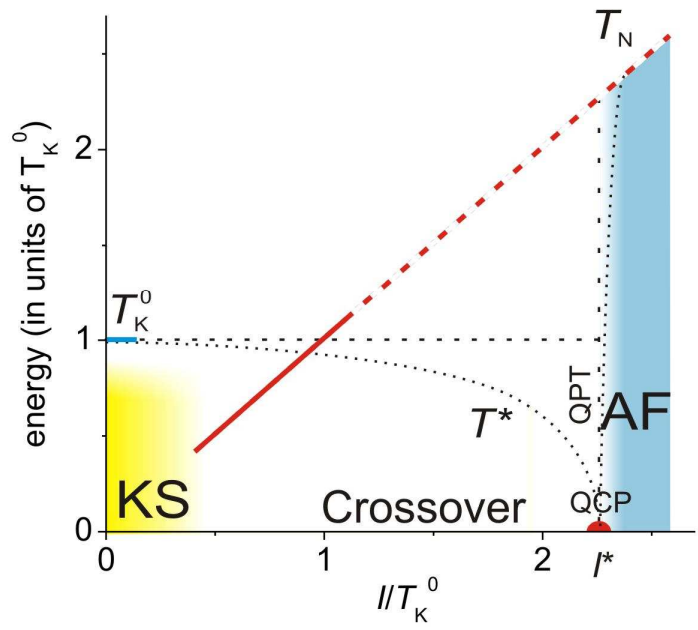

Fig. 5